\documentclass[]{pasj01}

\begin{document}
\Received{2015 March 27}
\Accepted{2015 June 10}
\Published{}
 
\title{High-resolution ALMA Observations of SDP.81. II. Molecular Clump Properties of a Lensed Submillimeter Galaxy at $z=3.042$}

\author{
Bunyo \textsc{Hatsukade}\altaffilmark{1*}
Yoichi \textsc{Tamura}\altaffilmark{2}
Daisuke \textsc{Iono}\altaffilmark{1,3}
Yuichi \textsc{Matsuda}\altaffilmark{1}
Masao \textsc{Hayashi}\altaffilmark{1}
and
Masamune \textsc{Oguri}\altaffilmark{4,5,6}
}
\altaffiltext{1}{National Astronomical Observatory of Japan, 2-21-1 Osawa, Mitaka, Tokyo 181-8588}
\email{bunyo.hatsukade@nao.ac.jp}
\altaffiltext{2}{Institute of Astronomy, University of Tokyo, 2-21-1 Osawa, Mitaka, Tokyo 181-0015}
\altaffiltext{3}{The Graduate University for Advanced Studies (SOKENDAI), 2-21-1 Osawa, Mitaka, Tokyo 181-8588}
\altaffiltext{4}{Research Center for the Early Universe, University of Tokyo, 7-3-1 Hongo, Bunkyo-ku, Tokyo 113-0033}
\altaffiltext{5}{Department of Physics, University of Tokyo, 7-3-1 Hongo, Bunkyo-ku, Tokyo 113-0033}
\altaffiltext{6}{Kavli Institute for the Physics and Mathematics of the Universe (Kavli IPMU, WPI), University of Tokyo, Chiba 277-8583}


\maketitle

\footnotemark[$*$] NAOJ Fellow

\begin{abstract}
We present spatially-resolved properties of molecular gas and dust in a gravitationally-lensed submillimeter galaxy H-ATLAS J090311.6$+$003906 (SDP.81) at $z=3.042$ revealed by the Atacama Large Millimeter/submillimeter Array (ALMA).
We identified 14 molecular clumps in the CO(5--4) line data.
The surface density of molecular gas ($\Sigma_{\rm H_2}$) and star-formation rate ($\Sigma_{\rm SFR}$) of the clumps are more than three orders of magnitude higher than those found in local spiral galaxies.
The clumps are placed in the `burst' sequence in the $\Sigma_{\rm H_2}$--$\Sigma_{\rm SFR}$ plane, suggesting that $z \sim 3$ molecular clumps follow the star-formation law derived for local starburst galaxies.
With our gravitational lens model, the positions in the source plane are derived for the molecular clumps, dust clumps, and stellar components identified in the {\sl Hubble Space Telescope} image.
The molecular and dust clumps are confined within a $\sim$2~kpc region, while the spatial extent of the stellar components is as large as $\sim$6~kpc and offset toward the west.
The molecular clumps have a systematic velocity gradient in the north-south direction, which may indicate a rotating gas disk.
One possible scenario is that the components of molecular gas, dust, and stars are distributed in a several-kpc scale rotating disk, and the stellar emission is heavily obscured by dust in the central star-forming region.
Alternatively, SDP.81 can be explained by a merging system, where dusty starbursts occur in the region where the two galaxies collide, surrounded by tidal features traced in the stellar components.
\end{abstract}

\section{Introduction}

Deep and wide-field submillimeter surveys have uncovered dust-obscured star-forming galaxies at high redshift (referred to as submillimeter galaxies; SMGs) (e.g., \cite{blai02}, for a review).
SMGs have intense star-forming activity with star-formation rates (SFRs) of a few 100--1000~$M_{\odot}$~yr$^{-1}$, which could be triggered by gas-rich galaxy mergers (e.g., \cite{tacc06}).
SMGs and local ultra-luminous infrared galaxies (ULIRGs) are known to be on the same relation on the molecular gas mass--far infrared (FIR) luminosity plane ($M_{\rm H_2}$--$L_{\rm FIR}$) or molecular gas surface density--SFR surface density plane ($\Sigma_{\rm H_2}$--$\Sigma_{\rm SFR}$), so-called `burst sequence', high above that of `normal' star-forming galaxies at local and $z \sim 1$--3 universe \citep{dadd10, genz10}.
Spatially resolved observations of molecular gas in SMGs are essential to investigate the star-forming properties.  However, only a handful of studies exists (e.g., \cite{swin10, shar13, rawl14, hodg15}) due to the limited spatial resolution and sensitivity of existing instruments.

Gravitational lensing is a powerful tool to probe the physical properties of distant galaxies in detail because of its magnification effect.
The gravitational lensing caused by a foreground massive galaxy or cluster of galaxies has been used to probe dusty star-forming galaxies at high-redshifts (e.g., \cite{smai97, smai02, negr10, viei13}).
\citet{swin10} discovered a $z = 2.3$ SMG (SMM~J2135$-$0102) which is lensed by a $z = 0.3$ foreground galaxy cluster with the magnification factor of $\mu = 32.5$,  allowing them to examine the properties of the SMG at 100~pc scale.

H-ATLAS J090311.6$+$003906 (SDP.81) is one of the brightest sources discovered in the {\sl Herschel} Astrophysical Terahertz Large Area Survey (HATLAS) \citep{eale10}.
Multi-wavelength follow-up observations have shown that SDP.81 is an SMG at $z = 3.042 \pm 0.001$ that is gravitationally lensed by a foreground massive elliptical galaxy at $z = 0.2999 \pm 0.0002$ \citep{negr10, negr14}.
Corrected for the gravitational lensing magnification factor of $\mu = 11$ \citep{buss13}, SDP.81 has an intrinsic FIR luminosity of $L_{\rm FIR} = 5 \times 10^{12}$~$L_{\odot}$ and a SFR of $\sim$500~$M_{\odot}$~yr$^{-1}$.
Recently, the Atacama Large Millimeter/submillimeter Array (ALMA) observed SDP.81 at 1--2~mm in the Science Verification observations with an angular resolution up to 0\farcs03 (which corresponds to $\sim$200~pc at $z=3.042$), offering an excellent opportunity to study the spatially-resolved properties of an SMG.

In this Letter, we present the properties of SDP.81 revealed by the high-angular resolution observations with ALMA.
Throughout the paper, we adopt a cosmology with $H_0=72$ km s$^{-1}$ Mpc$^{-1}$, $\Omega_{\rm{M}}=0.26$, and $\Omega_{\Lambda}=0.74$, and $1''$ corresponds to 7.78~kpc at $z=3.042$.

\section{Data}\label{sec:data}
 
\subsection{ALMA Data}\label{sec:alma}
 
We used the calibrated data set publicly available through the ALMA Science Portal.
The details of observations and data reduction are described in \citet{alma15a} and \citet{alma15b}.
SDP.81 was observed as part of Science Verification of the ALMA Long Baseline Campaign in October, 2014, at Band 4 (151~GHz, or 2.0~mm), Band 6 (236~GHz, or 1.3~mm), and Band 7 (290~GHz, or 1.0~mm) with the  baseline lengths of $\sim$5--15~km and with 31--36 antennas.

\begin{figure*}
\begin{center}
\includegraphics[width=\linewidth]{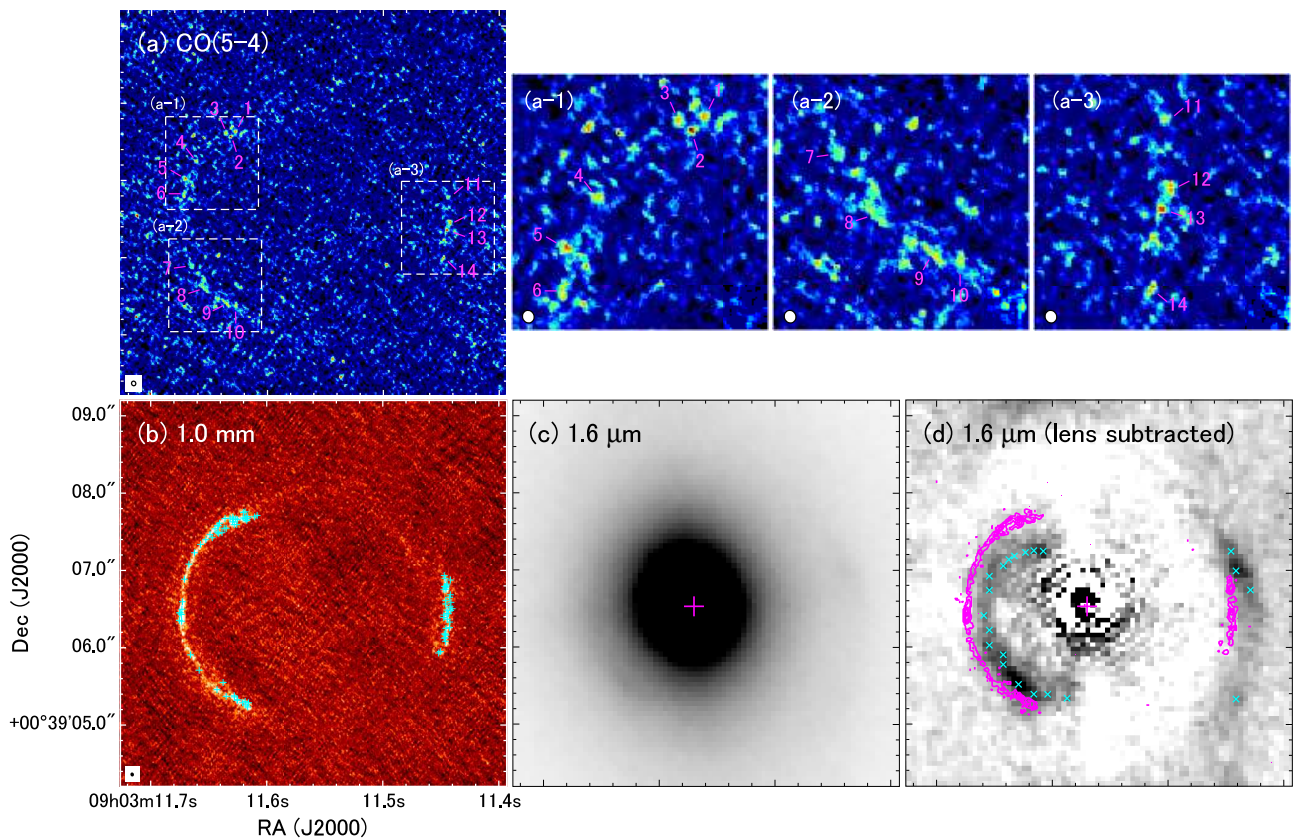}
\end{center}
\caption{
(a) ALMA CO(5--4) velocity-integrated intensity image.
Molecular clumps identified in this study are presented.
Synthesized beam is shown at the bottom left.
Magnified figures ($1\farcs2 \times 1\farcs2$) of (a-1), (a-2), and (a-3) are presented. 
(b) ALMA 1.0~mm continuum image.
Dust clumps identified in \citet{tamu15} are shown as plus marks.
Synthesized beam is shown at the bottom left.
(c) {\sl HST}/WFC3 1.6~$\mu$m.
Plus mark represents the position of the foreground lens galaxy detected in the ALMA 2.0~mm continuum image \citep{alma15b}.
(d) {\sl HST}/WFC3 1.6~$\mu$m lens-subtracted image with contours of the ALMA 1.0~mm image (from $+3$$\sigma$ with 2$\sigma$ step).
Plus mark represents the position of the foreground lens galaxy detected in the ALMA 2.0~mm continuum image .
Crosses represent the positions of stellar components which we used for deriving source-plane positions.
\label{fig:image}}
\end{figure*}

The data were reduced with the Common Astronomy Software Applications (CASA) \citep{mcmu07} package.
The maps were processed with the {\verb CLEAN } algorithm with the Briggs weighting (with {\verb robust } parameter of 1.0).
In this Letter, we used the Band 4 CO(5--4) line and the Band 7 1.0~mm continuum images without tapering the uv-data at long baselines.
The achieved synthesized beamsize (full-width at half maximum) is $0\farcs060 \times 0\farcs054$ and $0\farcs031 \times 0\farcs023$ for the CO(5--4) and 1.0~mm images, respectively.  
The images are shown in Figure~\ref{fig:image}.
Two arc structures in east and west caused by gravitational lensing are clearly seen.

\subsection{{\sl HST} Data}\label{sec:hst}
 
We made use of the archival data of {\sl HST}/Wide Field Camera 3 (WFC3) taken by \citet{negr14}.
We retrieved the calibrated, flat-fielded individual exposure frames of WFC3 with the F160W filter ($H$-band) from the archive, and then coadded them by ourselves using the {\sc iraf} {\sc MultiDrizzle} package.
The final pixel scale was set to 0\farcs064  to improve the resolution and quality of the image.
We calibrated the coordinates of the image using the catalog data from the Data Release 12 of the Sloan Digital Sky Survey (SDSS)\footnotemark.
We used 46 SDSS sources located in the image to derive the astrometric fit using {\sc iraf}.
The rms error to the fit is 0\farcs085 in right ascension and 0\farcs100 in declination.
We checked that the coordinates of the foreground lens galaxy in the SDSS catalog match the coordinates of the continuum emission detected in Band 4, 6, and 7 \citep{alma15b} within 0\farcs015.
The offset between the center of the foreground lens and the ALMA continuum emission is smaller than the ALMA beamsize, showing the consistency of astrometry between the {\sl HST} and ALMA images (Figure~\ref{fig:image}c).

We used the software {\sc galfit} \citep{peng02} to deblend the background lensed source from the foreground lens galaxy.
According to \citet{negr14}, two S\'{e}rsic profiles were simultaneously fitted to the foreground lens galaxy to cleanly subtract it.
As a result, the brighter component of the foreground lens galaxy has the S\'{e}rsic index of $n=4.42$ (elliptical), while the fainter component has the index of $n=1.03$ (exponential).
The profiles are consistent with those obtained by \citet{negr14}.
The stellar arc structure seen in the lens-subtracted image shown in Figure~\ref{fig:image}d is  similar to the arcs seen in the ALMA CO line and continuum images, but they are clearly offset toward the west.

\footnotetext{http://skyserver.sdss.org/dr12/}

\section{Lens Model}\label{sec:lens}
 
The analysis of gravitational lensing was conducted by \citet{tamu15} using the {\sc glafic} software package \citep{ogur10}.
The lens model was constructed by identifying the knots which were common in all three continuum images
at 1.0, 1.3 and 2.0~mm.
The lensed images are explained by the best-fit mass model with a mass profile of an isothermal ellipsoid with a 400-pc core.
The core radius of the isothermal ellipsoid agrees well with that of the stellar light in the {\sl HST}/WFC3 image,
and the position of the predicted mass centroid is consistent with that of the emission detected in the ALMA continuum images.
\citet{tamu15} found that the continuum emission has a power-law spectral index of $-0.64 \pm 0.1$ from 1.0~mm to 20~cm, and it is likely that the non-thermal emission arises from an active galactic nucleus of the lens galaxy.
This lens model is used to compute the positions on the source plane (Section~\ref{sec:sdp81}).

\section{Molecular Clumps}\label{sec:clump}
 
\subsection{Clump Identification}\label{sec:id}
 
We identified molecular clumps by using the line cube (spectral resolution of 21 km~s$^{-1}$) and the velocity-integrated intensity map of CO(5--4).
We first searched for the peak positions with $\ge$3.5$\sigma$ in the velocity-integrated intensity map created by including pixels above 1.5$\sigma$ in the line cube.
We then used the {\sc clumpfind} package \citep{will94} to search for clumps in the line cube.
We identified 14 molecular clumps along the arcs detected by both methods (Figure~\ref{fig:image}a).
The positions, magnification factors, and mean velocities are shown in Table~\ref{tab:property}.
We caution here that the gravitational lensing can possibly produce two (or multiple) clumps in the image plane 
that are in fact the same clump in the source plane.   
However, because of the low SN, it may be difficult to identify all of the multiples in the CO(5--4) image.
We use the positions in image plane and source plane, and the mean velocities, as a guide to identify multiple 
images, and we suggest that clumps 1 and 13 and clumps 4 and 5 are possibly counter images of the same source.

If the clumps are spatially resolved with the ALMA beam, the transverse direction should be more extended than the radial direction.
We found that some of the clumps do not appear to be extended in the transverse direction, and it is likely that they are unresolved.
In order to measure the size of the clumps, higher angular resolution observations are needed.

The magnification of the four clumps close to the critical curve (ID 4, 5, 7, and 8) is large ($>$100) and varies significantly within each clump.
It is possible that they have properties with different physical scales compared to the other clumps, and we exclude these clumps in the discussion of star formation properties. 
Except for the four clumps, the intrinsic spatial scale is roughly a few 100~pc in the source plane after correcting for the magnification factor and assuming that the clumps are spatially resolved.

\subsection{Surface densities of molecular gas and SFR}\label{sec:properties}
 
We derived the surface density of the molecular gas and the SFR in order to investigate the intrinsic properties of SDP.81. 
We assume that there is no significant difference in the distribution of CO and dust emission within the measured region for each clump, and thus the effect of differential magnification (e.g., \cite{blai99, serj12, heza12}) is negligible.
Because the distribution of molecular clouds or star-forming regions smaller than the beamsize is unknown, we assume a filling factor of unity in this study.
We note that the filling factor might be less than unity if they are not spatially resolved.

We used regions above 2$\sigma$ in the CO(5--4) velocity-integrated map for conducting photometry and estimating the physical quantities of each clump.
The photometry of dust continuum was conducted on the 1.0~mm image in the same regions used for the CO photometry. 
The CO luminosity of each clump was calculated from $L'_{\rm CO} = 3.25 \times 10^7 S_{\rm CO}\Delta v \nu_{\rm obs}^{-2} D_{\rm L}^2 (1+z)^{-3}$ \citep{solo05}, where $S_{\rm CO}\Delta v$ is the velocity-integrated intensity in Jy~km~s$^{-1}$, $\nu_{\rm obs}$ is the observed line frequency in GHz, and $D_{\rm L}$ is the luminosity distance in Mpc.
The error of $S_{\rm CO}\Delta v$ is calculated by using the equation of \citet{hain04}.
We used a CO line ratio of CO(5--4)/CO(1--0) $= 0.3$ \citep{alma15b}.
The molecular gas mass is derived from $M_{\rm H_2} = \alpha_{\rm CO} L'_{\rm CO(1-0)}$, where $\alpha_{\rm CO}$ is the CO-to-molecular gas mass conversion factor including He mass.
We adopted a conversion factor of $\alpha_{\rm CO} = 0.8$ $M_{\odot}$~(K~km~s$^{-1}$~pc$^2$)$^{-1}$, the standard value for ULIRGs \citep{down98}.
The FIR luminosity and dust mass were derived from $L_{\rm FIR} = 4\pi M_{\rm dust} \int_0^{\infty} \kappa_d(\nu_{\rm{rest}})B(\nu_{\rm{rest}}, T_{\rm dust}) d\nu$, and from $M_{\rm dust}=S_{\rm{obs}}D_L^2/[(1+z)\kappa_d(\nu_{\rm{rest}})B(\nu_{\rm{rest}}, T_{\rm dust})]$ \citep{debr03},
where $\kappa_d(\nu_{\rm{rest}})$ is the dust mass absorption coefficient, $\nu_{\rm{rest}}$ is the rest-frame frequency, $T_{\rm dust}$ is the dust temperature, $B(\nu_{\rm{rest}}, T_{\rm dust})$ is the Planck blackbody function, and $S_{\rm{obs}}$ is the observed flux density.
We assume that the absorption coefficient varies as $\kappa_d \propto \nu^{\beta}$ and the emissivity index lies between 1 and 2 (e.g., \cite{hild83}), and adopt $\kappa_d(125\ \mu m) = 2.64 \pm 0.29$~m$^2$~kg$^{-1}$ \citep{dunn03} and $\beta = 1.5$.
The dust temperature of $T_{\rm dust} = 39.3$~K was derived by \citet{negr14} for the integrated value of SDP.81, which is comparable to high-redshift ULIRGs and SMGs (e.g., \cite{kova06, syme13}). 
We adopted the dust temperature for each clump. 
The SFR was derived from SFR $= 1.72 \times 10^{-10} L_{\rm FIR}$ \citep{kenn98a}.
The derived surface densities of molecular gas and SFR are shown in Table~\ref{tab:property}.

\begin{table*}
\tbl{Properties of molecular clumps. \label{tab:property}}{
\scriptsize
\begin{tabular}{ccccccccccc}
\hline
ID & RA$_{\rm image}$$^a$ & Dec$_{\rm image}$$^a$ & RA$_{\rm source}$$^b$ & Dec$_{\rm source}$$^b$
& $\mu_{\rm mean}$$^c$ & $\mu_{\rm min}$$^c$ & $\mu_{\rm max}$$^c$ & $v_{\rm mean}$$^d$ & $\Sigma_{\rm H_2}$$^e$ & $\Sigma_{\rm SFR}$$^f$ \\
   & & & & & & & & (km~s$^{-1}$) & ($10^4 M_{\odot}$~pc$^{-2}$) & ($10^2 M_{\odot}$~yr$^{-1}$~kpc$^{-2}$) \\
\hline
 1 & 09:03:11.628 & $+$00:39:7.63 & 09:03:11.558 & $+$00:39:06.47 &      4 &  3.6 &    4.7 & $ -94  $ & $1.11 \pm 0.33$ & $3.80 \pm 0.19$ \\
 2 & 09:03:11.631 & $+$00:39:7.56 & 09:03:11.556 & $+$00:39:06.47 &      4 &  3.7 &    4.7 & $-165  $ & $1.19 \pm 0.36$ & $1.60 \pm 0.20$ \\
 3 & 09:03:11.636 & $+$00:39:7.62 & 09:03:11.559 & $+$00:39:06.53 &      6 &  4.9 &    6.6 & $ -78  $ & $1.13 \pm 0.35$ & $1.63 \pm 0.20$ \\
 4 & 09:03:11.662 & $+$00:39:7.25 & 09:03:11.560 & $+$00:39:06.62 & $>$100 & 17   & $>$100 & $ 118  $ & $1.09 \pm 0.27$ & $1.72 \pm 0.15$ \\
 5 & 09:03:11.671 & $+$00:39:7.02 & 09:03:11.561 & $+$00:39:06.62 & $>$100 & 14   & $>$100 & $ 107  $ & $1.04 \pm 0.19$ & $1.58 \pm 0.11$ \\
 6 & 09:03:11.672 & $+$00:39:6.78 & 09:03:11.559 & $+$00:39:06.60 &     26 & 15   &     67 & $  92  $ & $1.12 \pm 0.28$ & $1.79 \pm 0.16$ \\
 7 & 09:03:11.661 & $+$00:39:5.79 & 09:03:11.560 & $+$00:39:06.53 & $>$100 & 17   & $>$100 & $  20  $ & $1.10 \pm 0.36$ & $1.53 \pm 0.20$ \\
 8 & 09:03:11.653 & $+$00:39:5.57 & 09:03:11.562 & $+$00:39:06.50 & $>$100 & 12   & $>$100 & $ -21  $ & $1.05 \pm 0.16$ & $1.19 \pm 0.09$ \\
 9 & 09:03:11.635 & $+$00:39:5.39 & 09:03:11.560 & $+$00:39:06.54 &     14 &  9.7 &   21   & $  40  $ & $1.12 \pm 0.25$ & $1.39 \pm 0.14$ \\
10 & 09:03:11.627 & $+$00:39:5.39 & 09:03:11.558 & $+$00:39:06.60 &      8 &  6.5 &   11   & $  29  $ & $1.08 \pm 0.31$ & $2.91 \pm 0.17$ \\
11 & 09:03:11.444 & $+$00:39:6.77 & 09:03:11.558 & $+$00:39:06.57 &      4 &  3.9 &    4.4 & $  92  $ & $1.04 \pm 0.35$ & $3.09 \pm 0.20$ \\
12 & 09:03:11.442 & $+$00:39:6.44 & 09:03:11.557 & $+$00:39:06.49 &      4 &  3.8 &    4.3 & $   0.9$ & $1.15 \pm 0.27$ & $2.94 \pm 0.15$ \\
13 & 09:03:11.445 & $+$00:39:6.36 & 09:03:11.559 & $+$00:39:06.47 &      4 &  4.0 &    4.4 & $ -90  $ & $1.16 \pm 0.31$ & $2.08 \pm 0.18$ \\
14 & 09:03:11.448 & $+$00:39:5.99 & 09:03:11.558 & $+$00:39:06.37 &      4 &  4.1 &    4.6 & $ -98  $ & $1.13 \pm 0.32$ & $1.39 \pm 0.18$ \\
\hline
\end{tabular}}
\begin{tabnote}
Note: ID 1 and 13, and ID 4 and 5 are possibly counter images of the same source.
$^a$ Right ascension and declination of the ALMA peak positions in the image plane (J2000);
$^b$ Right ascension and declination in the source plane (J2000);
$^c$ Mean, minimum, and maximum magnification factors measured within the region of each clump;
$^d$ Mean velocity measured within the region of each clump;
$^e$ Molecular gas surface density;
$^f$ SFR surface density.
\end{tabnote}
\end{table*}

\section{Discussion}\label{sec:discussion}
 
\subsection{Star-formation Properties}\label{sec:sk}
 
\begin{figure}
\begin{center}
\includegraphics[width=\linewidth]{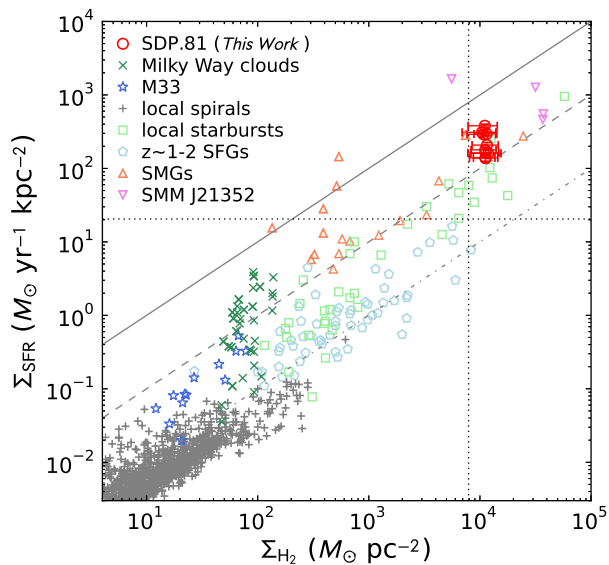}
\end{center}
\caption{
Molecular gas surface density--SFR surface density plot.
The molecular clumps of SDP.81 are plotted as red circles.
The vertical and horizontal dotted line represents the 2$\sigma$ detection limit on the surface density of molecular gas and SFR, respectively, for a clump with a source size comparable to the beamsize. 
For comparison we also plot data points taken from literature:
Milky Way clouds \citep{heid10, evan14},
giant {\sc Hii} regions in M33 \citep{miur14},
local disk galaxies \citep{bigi10, kenn98b},
local starbursts \citep{kenn98b},
$z \sim 1$--2 star-forming galaxies \citep{tacc08},
SMGs \citep{both10},
and a spatially-resolved lensed SMG SMM~J21352 \citep{thom15}.
The solid, dashed, and dot-dashed lines represent the gas depletion time scale of 10~Myr, 100~Myr, and 1~Gyr, respectively.
If the filling factor is less than unity, the data points of SDP.81 shift to the upper-right direction.
\label{fig:sk}}
\end{figure}

The surface density of molecular gas and SFR for the molecular clumps are plotted in Figure~\ref{fig:sk}.
The magnification varies with each clump, and a clump with a larger magnification traces an intrinsically fainter one.
We reiterate that the magnification correction is not necessary when discussing the surface density of molecular gas and SFR, because the gravitational lensing conserves the surface brightness for the identified clumps.
The molecular clumps of SDP.81 traced in this study have surface densities that are more than three orders of magnitude higher than those of local spiral galaxies, and they are similar to those of the $\sim$100~pc scale clumps of the spatially-resolved SMG SMM~J21352 \citep{thom15}.
We note that the four clumps with high magnification ($>$100; Section~\ref{sec:id}) which are not presented in Figure~\ref{fig:sk} share the same region with the other clumps. 
The SFR density ranges between 100--300 $M_{\odot}$~yr$^{-1}$~kpc$^{-2}$, which is lower than the Eddington-limited star formation (maximum starburst; \cite{elme99}) \citep{ryba15}.
It is known that local star-forming galaxies follow a relation in the $\Sigma_{\rm H_2}$--$\Sigma_{\rm SFR}$ plane (Schmidt--Kennicutt relation; \cite{schm59, kenn98a}), and it is extended to include higher redshift sources (e.g., \cite{genz10}).  The clumps of SDP.81 are located in the `burst' sequence \citep{dadd10} with a gas depletion time scale of $\sim$100~Myr, suggesting active star-forming activity in the clumps.
If we adopt the Galactic conversion factor ($\alpha_{\rm CO} = 4.3 M_{\odot}$~(K~km~s$^{-1}$~pc$^2$)$^{-1}$; \cite{bola13}), these clumps are located on the sequence of normal star-forming galaxies.
Note that if the filling factor is less than unity, the data points in Figure~\ref{fig:sk} will shift to the upper-right direction.
Regardless of the exact location of these clumps in the $\Sigma_{\rm H_2}$--$\Sigma_{\rm SFR}$ plane,
the derived $\Sigma_{\rm SFR}/\Sigma_{\rm H_2}$ ratios are consistent with the values found in the Milky Way clumps (e.g., \cite{heid10, evan14}), suggesting that the star-formation law can be applied to molecular clumps at $z \sim 3$.

\subsection{The Nature of SDP.81}\label{sec:sdp81}
 
\begin{figure}
\begin{center}
\includegraphics[width=.8\linewidth]{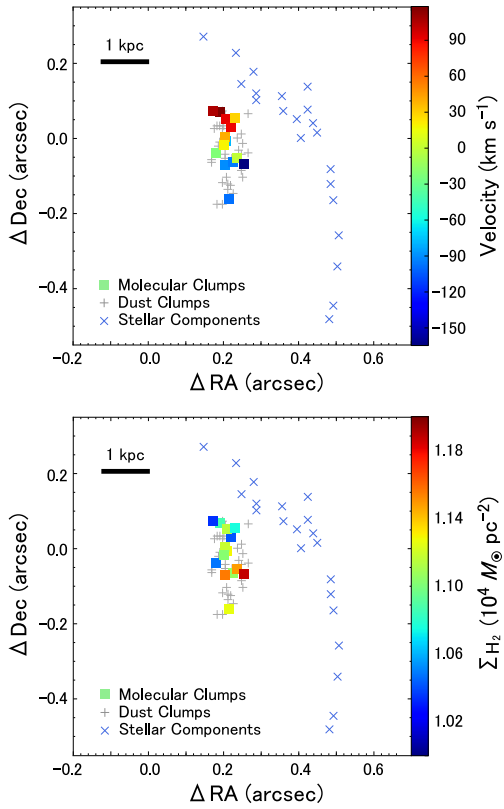}
\end{center}
\caption{
Positions in the source plane for the molecular clumps, the dust clumps, and the stellar components marked in Figure~\ref{fig:image}.
Velocity (top) and surface density (bottom) of the molecular clumps are indicated by color scale.
\label{fig:source-plane}}
\end{figure}

By using our lens model \citep{tamu15} and the {\sc glafic} software, we determined the positions of molecular clumps on the source plane (Figure~\ref{fig:source-plane}).
The clumps are distributed over a 2~kpc area, and they overlap with the dust clumps identified in \citet{tamu15}.
We also derived the source-plane positions for the brighter points of the stellar components
shown as crosses in Figure~\ref{fig:image}d.
The points are selected so that the relative correspondence of the stellar components between the image plane and the source plane is easily visualized.
The stellar components traced in this study are distributed over $\sim$6~kpc in the NE-SW direction, which is consistent with the result of \citet{dye14}, and offset from the molecular gas and dust clumps toward the west.
Since the lensing model of \citet{tamu15} includes the contribution from the stellar emission detected in the {\sl HST} image, we argue that the extension of the stellar emission is real and not due to the lens model being based on the clumps of ALMA images.
We note that, since the main aim of this study is to check the positional correlation between molecular gas/dust and stars, we only computed the position of the components marked in Figure~\ref{fig:image}d, and the spatial extent of each component is not taken into account.

The mean velocity of each molecular clump measured in the image plane is shown in color in  
Figure~\ref{fig:source-plane} (top). A velocity gradient with a peak-to-peak value of $\sim$300~km~s$^{-1}$ in the north-south direction can be seen, which could indicate a rotating gas disk.
In Figure~\ref{fig:source-plane} (bottom), the molecular gas surface density is shown in color scale.
The molecular gas surface density appears to be higher in the central part of the molecular/dust `diskf and decreasing toward the edge, although the uncertainty is large.

What is the nature of SDP.81?
One possible scenario is that the components of molecular gas, dust, and stars are distributed in a several-kpc scale rotating disk, and the stellar emission is heavily obscured by dust in the central star-forming region.
This is qualitatively consistent with the $z=4.05$ SMG GN20, where an offset of rest-frame UV-emitting region by 4~kpc from a region with CO and dust emission has been found \citep{iono06, hodg15}.
Alternatively, SDP.81 may be a merging system, where dusty starbursts occur in the interface between the two colliding galaxies while the tidal features are widely extended.  A similar large scale velocity gradient is seen in a recent ALMA CO(3--2) image of the Antennae Galaxies (\cite{whit14}).
The CO(5--4) line we used in this study has a relatively high critical density ($10^{5}$~cm$^{-3}$) and it may not trace the gas of the entire disk.  Future observations of the lower transition CO lines will allow us to study the detailed distribution and kinematics of the diffuse gas, providing further insights to the exact nature of SDP.81.

\begin{ack}
We are grateful to Rie Miura and Alasdair Thomson for providing data.
We thank the referee for helpful comments and suggestions.
BH is supported by JSPS KAKENHI Grant Number 15K17616. 
YT is supported by KAKENHI (No. 25103503).
DI is supported by the 2015 Inamori Research Grants Program.
YM is supported by KAKENHI (No. 20647268).
MH is supported by Research Fellowship for Young Scientists from the Japan Society of the Promotion of Science (JSPS).
MO is supported in part by World Premier International Research Center Initiative (WPI Initiative), MEXT, Japan and Grant-in-Aid for Scientific Research from the JSPS (26800093).
This paper makes use of the following ALMA data: ADS/JAO.ALMA\#2011.0.00016.SV.
ALMA is a partnership of ESO (representing its member states), NSF (USA) and NINS (Japan), together with NRC (Canada), NSC and ASIAA (Taiwan), and KASI (Republic of Korea), in cooperation with the Republic of Chile. The Joint ALMA Observatory is operated by ESO, AUI/NRAO and NAOJ.
Data analysis were in part carried out on common use data analysis computer system at the Astronomy Data Center, ADC, of the National Astronomical Observatory of Japan.
\end{ack}


\end{document}